\documentclass[preprint,pteplogo]{ptephy}

\preprintnumber{XXXX-XXXX} 

\usepackage{amsmath} 
\usepackage{amssymb}
\usepackage{bm}

\begin{document}

\title{Concentration of the velocity distribution of pulsed neutron beams}

\author{\name{Masaaki Kitaguchi}{1,3}, \name{Yoshihisa Iwashita}{2}, and \name{Hirohiko M. Shimizu}{3}}

\address{\affil{1}{Center for Experimental Studies, Nagoya University}
\affil{2}{Institute for Chemical Research, Kyoto University}
\affil{3}{Department of Physics, Nagoya University}
\email{kitaguchi@phi.phys.nagoya-u.ac.jp}}

\begin{abstract}%
The velocity of neutrons from a pulsed neutron source is well-defined as a function of their arrival time.
Electromagnetic neutron accelerator/decelerator synchronized with the neutron time-of-flight is capable of selectively changing the neutron velocity and concentrating the velocity distribution.
Possible enhancement of the neutron intensity at a specific neutron velocity by orders of magnitude is discussed together with an experimental design.
\end{abstract}

\subjectindex{xxxx, xxx}

\maketitle


\section{Introduction}
Increase of the production efficiency of ultra-cold neutrons (UCNs) in the energy range of below about $250$ neV is intensively attempted for the improvement of the experimental accuracies of neutron properties such as the neutron electric dipole moment (nEDM), $\beta$-decay lifetime and searches for exotic interactions with neutrons~\cite{refEDM, refLife}.
The UCN storage density is the crucial parameter especially for the suppression of the systematic errors in the measurement of the nEDM.
The combination of an accelerator-based spallation neutron source and a superthermal down-conversion from the very-cold region into the ultracold region introduced the flexibility in the design of UCN sources and has been employed for new generation UCN sources~\cite{refUCN}.
Solid deuterium has been utilized at currently operational UCN sources at Los Alamos National Laboratory~\cite{refANL} and Paul Scherrer Institut~\cite{refPSI}.

Application of the inelastic neutron scattering due to phonon excitations in superfluid helium is expected to further improve the UCN density since the UCN loss due to the absorption and up-scattering can be sufficiently suppressed below the temperature of about $1$ K although its down-conversion rate is relatively smaller than that of the solid deuterium.
Superfluid helium is adopted as UCN sources under construction at TRIUMF~\cite{refTRIUMF}, in preparation at the SNS~\cite{refSNS}, under design study at the PNPI~\cite{refPNPI}.
The down-conversion into the UCN region is dominated by the single phonon exitation induced by very-cold neutrons with the velocity $v_n$ of around $447$ m s$^{-1}$ (the wavelength of around $0.89$ nm).
Therefore, UCN yield is approximately proportional to the neutron flux at $v_n \sim 447 ({\rm m}\,{\rm s}^{-1})$ and the TRIUMF and PNPI sources are designed to maximize the coupling efficiency among the neutron production target, cold moderator and the superfluid helium converter.

The SNS source is designed to accept a pulsed cold neutron beam into a superfluid helium.
Only neutrons with the velocity of around $447$m/s are contributing to the UCN yield while other neutrons simply transmit the superfluid helium.

In this paper, we discuss an application of neutron accelerator/decelerator, that was demonstrated to achieve the UCN time focusing~\cite{refRebuncher}, to concentrate neutrons into the $v_n \sim 447 ({\rm m}\,{\rm s}^{-1})$ region by selectively decelerate faster neutrons and accelerate slower neutrons on 
the beam transportation 
to the superfluid helium with a series of accelerator/decelerator units synchronized with the neutron time-of-flight.

\section{Concept of neutron velocity concentrator}

\begin{figure}[th]
\centering\includegraphics[width=10cm]{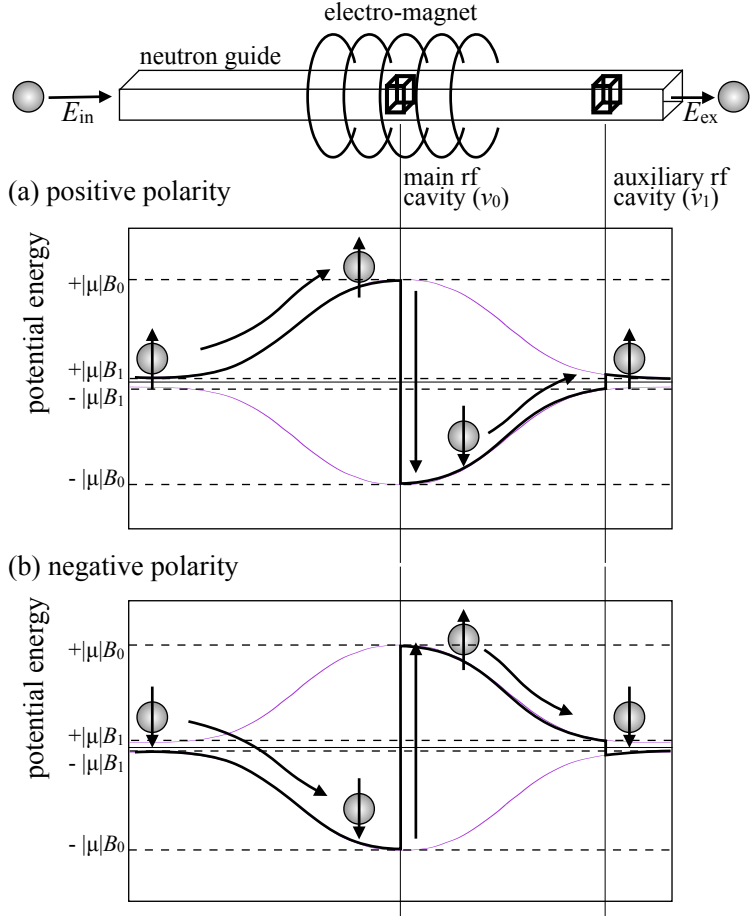}
\caption{Schematic view of a spin flipper unit. 
(a) A neutron with positive polarity is decelerated.
(b) A neutron with negative polarity is accelerated.
}
\label{fig_pot}
\end{figure}

We denote the longitudinal flight path length as $z$.
We consider the case where neutrons are transported through a magnetic field using a neutron guide and the transverse velocity is sufficiently smaller than the longitudinal velocity, where the neutron motion can be described as a one-dimensional problem along the $z$-axis as schematically illustrated in Fig.~\ref{fig_pot}.
We assume that the magnetic field is sufficiently strong so that the polarity of the neutron spin about the local magnetic field is adiabatically transported.
In general, the neutron polarity under a magnetic field $\bm{B}$ can be flipped by applying an rf-field with the frequency of $\nu=2| \mu \bm{B}|/h$, where $\mu$ is the neutron magnetic moment.
If the incident neutron spin polarity is positive, the relation between the kinetic energy of the incident neutron ($E_{\rm in}$) and that of the exit neutron ($E_{\rm ex}$) is
$
E_{\rm ex} = E_{\rm in} - h\nu,
$
and the neutron is decelerated by $h\nu$.
On the other hand, negative polarity neutron is accelerated by $h\nu$.

Here we consider the case where a main rf-cavity is installed in a strong-field region on the neutron beam path and an auxiliary rf-cavity in a weak-field region as shown in Fig.~\ref{fig_pot}.
The frequency of the main rf-cavity ($\nu_0$) is significantly larger than that of the auxiliary rf-cavity ($\nu_1$).
We refer to this device as a ``spin flipper unit'' below.
The spin flipper unit decelerates positive polarity neutrons by $h\nu_0$ and it accelerates negative polarity neutrons by $h\nu_0$.
The spin polarity is flipped to the original polarity in the auxiliary rf-cavity.

The rf-power is applied for a certain duration time earlier than $z_i/v_{\rm f}$ when neutrons faster than $v_{\rm f}$ arrive $i$-th spin flipper unit.
Positive polarity neutrons with the velocity $v > v_{\rm f}$ are decelerated and negative polarity ones accelerated.

\begin{figure}[th]
\centering\includegraphics[width=13cm]{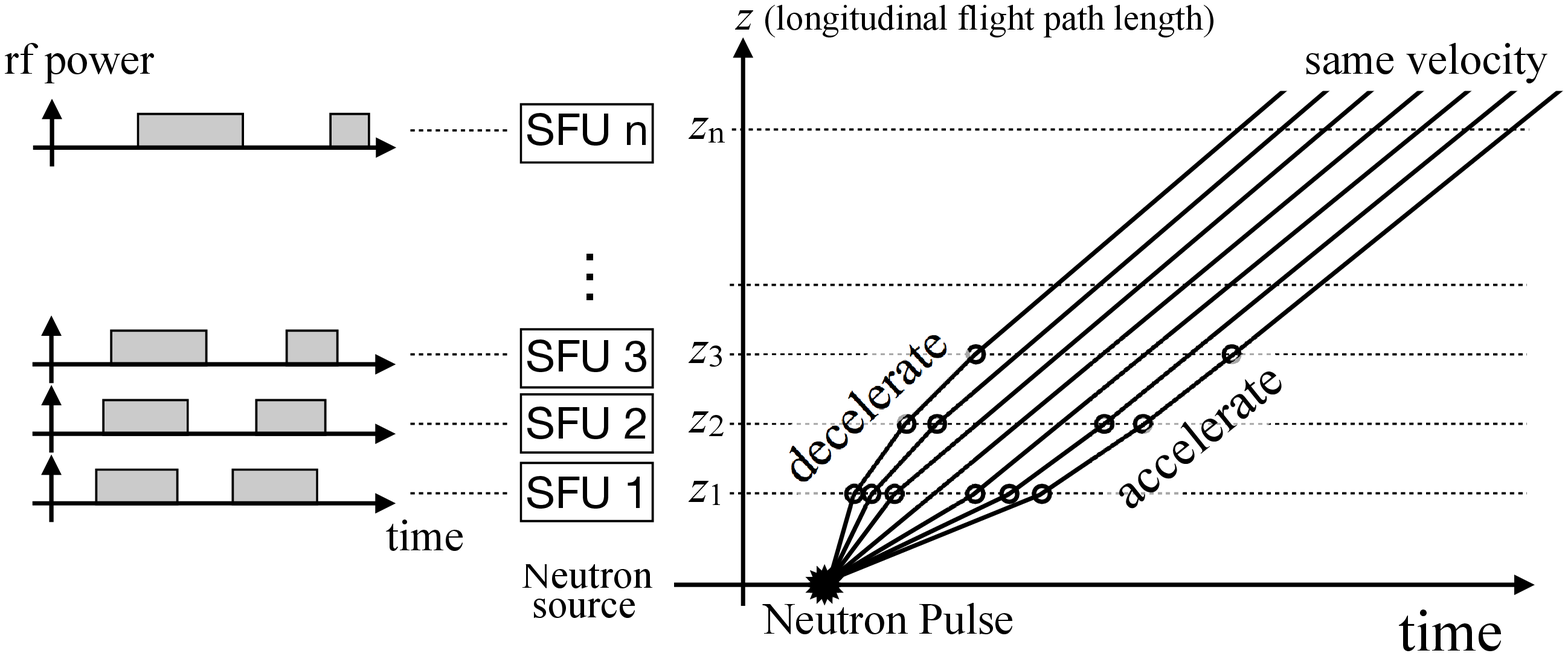}
\caption{Schematic drawing of the neutron velocity concentrator and the timing chart of rf-power application.}
\label{fig_position}
\end{figure}

Figure~\ref{fig_position} illustrates a series of identical spin flipper units placed with an equal spacing to which we refer as the ``neutron velocity concentrator''.
We denote the distance between the neutron source and the main rf-cavity, at which spin polarity is flipped, of $i$-th spin flipper unit as $z_i$.
We consider to concentrate neutron velocity distribution around a specific velocity which is referred to as the target velocity $v_{\rm target}$ below.
The rf-power of $i$-th spin flipper unit is turned off at the time of $t=z_i/v_{\rm target}$ when neutrons with the velocity $v_{\rm target}$ arrive.

\begin{figure}[th]
\centering\includegraphics[width=15cm]{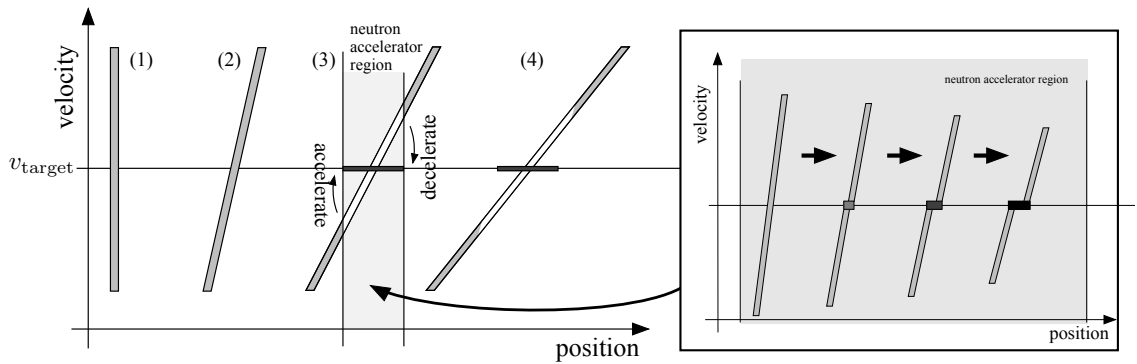}
\caption{Time evolution of the phase space distribution of neutrons. 
(1) Neutrons are generated as a pulse. 
(2) Neutrons spread spatially during transport. 
(3) Neutrons are accelerated or decelerated properly. 
(4) Velocity distribution in concentrated into the vicinity of the target velocity.
Right box shows that of a neutron bunch in the accelerator region. 
This figure illustrates the behavior of polarized neutrons.
}
\label{fig_phasespace}
\end{figure}

If the first spin flipper unit (SFU 1) is on at the time $z_1/v$ and the $\nu_0$ satisfies Eq.~\ref{eq:single}, the velocity of positive polarity component is changed to $v_{\rm target}$ in the SFU 1.
\begin{equation}
	\frac{mv_{\rm target}^2}{2} = \frac{mv^2}{2}-h\nu_0
	\label{eq:single}
\end{equation}
Much faster neutrons can be decelerated in the vicinity of $v_{\rm target}$ by the successive deceleration through SFU 1, SFU 2, \ldots.
In the same way, slower neutrons can be accelerated into the vicinity of $v_{\rm target}$.
Selecting the timing of the rf-power application, 
neutron velocity distribution can be concentrated into the vicinity of $v_{\rm target}$ as shown in Fig.~\ref{fig_phasespace}.

We note that the final polarity of concentrated neutrons can be arranged to be positive by turning off the auxiliary rf power of the last active spin flipper unit. 

\section{A design of neutron velocity concentrator for superthermal ultracold source with superfluid helium}\label{sec_simulation}

\begin{figure}[th]
\centering\includegraphics[width=8cm]{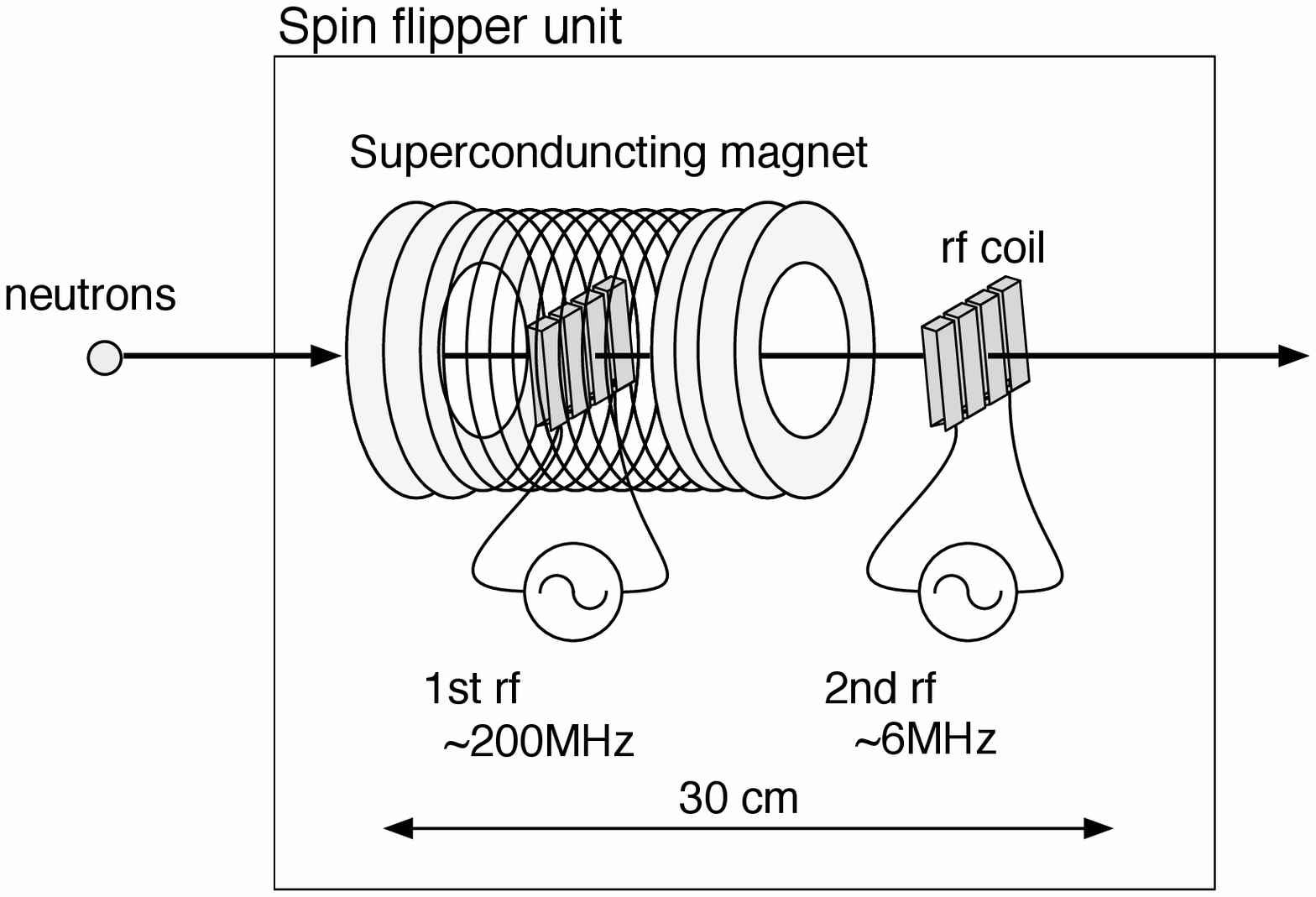}
\centering\includegraphics[width=6cm]{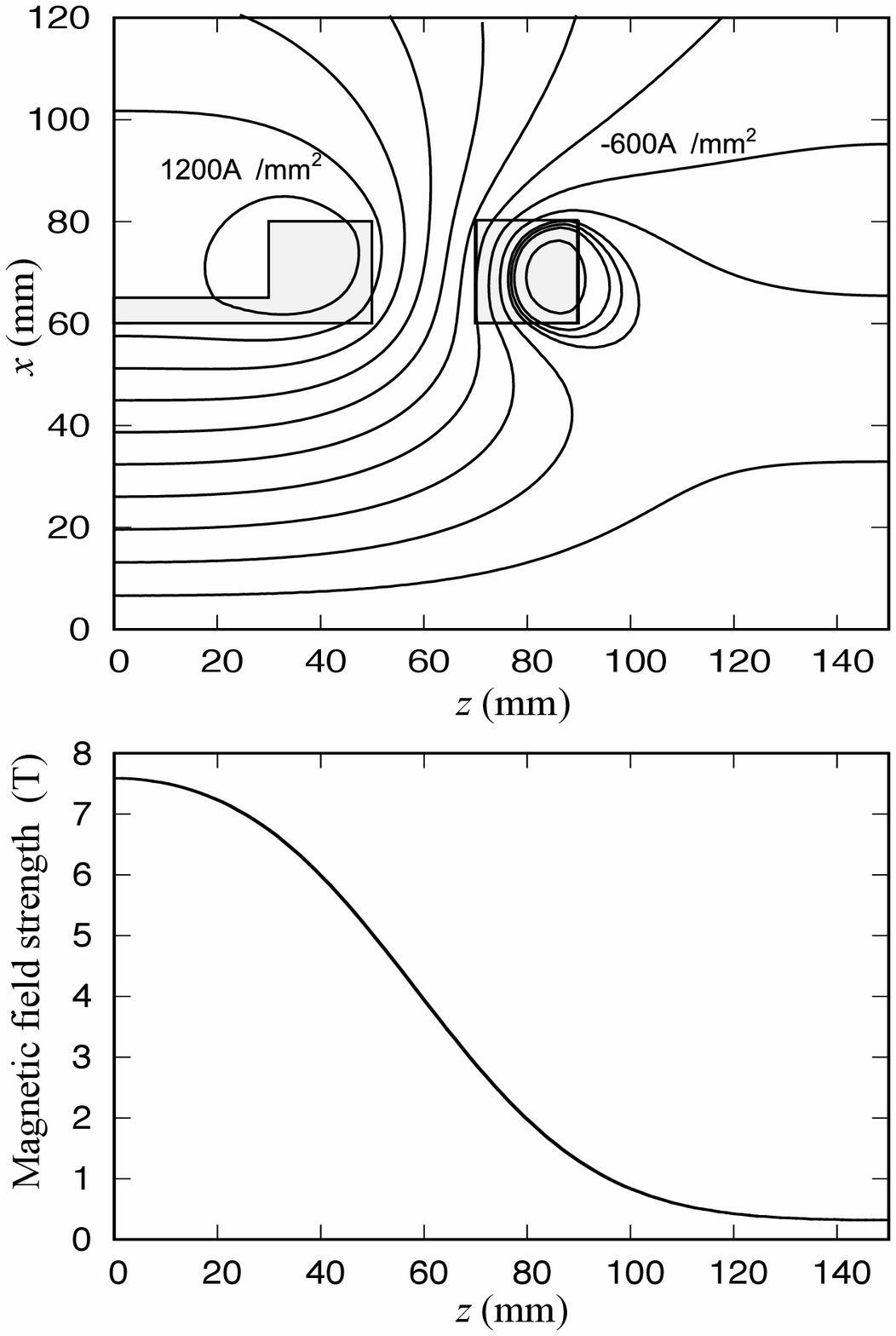}
\caption{Left: Conceptual design of spin flipper unit.
Right-Top: A model of superconducting solenoid magnet with 
the symmetry about $z=0$ plane and the symmetry around $z$ axis. 
Gray areas represent superconducting solenoid coil. 
Generated flux lines are also shown. 
Curves represent lines of magnetic force. 
Right-Bottom: Calculated magnetic field strength. }
\label{fig_scm}
\end{figure}

We discuss a practical design for concentrating the neutron velocity distribution into the vicinity of $v_{\rm n} \sim 447 \,({\rm m}\,{\rm s}^{-1})$ where the down-conversion into the ultracold region via phonon excitations in superfluid helium takes place.
We employ a superconducting compact solenoid as the electromagnet of each spin flipper unit.
The numerical calculation tells us that the field strength can be $7.5$ T and the length of the spin flipper unit can be $30$ cm as shown in Fig.\ref{fig_scm}.
Corresponding rf frequency for the spin flip is about $200$ MHz and resulting neutron energy change is about $0.9$ $\mu$eV.

\begin{figure}[th]
\centering\includegraphics[width=13cm]{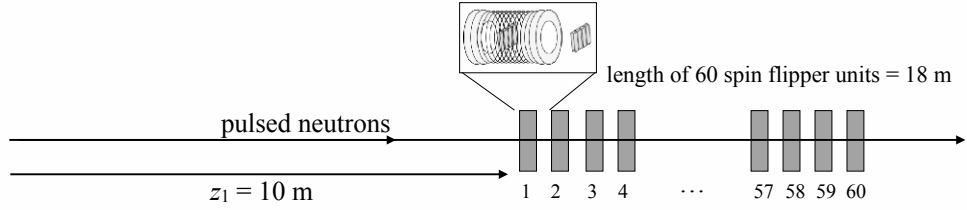}
\caption{Beamline setup for simulation. 
Sixty flipper units were set on the beamline at the position of 10 m from the source.}
\label{fig_beamline}
\end{figure}

We employ a series of sixty identical spin flipper units as shown in Fig.\ref{fig_beamline}.
We simulated the transport of pulsed neutrons by using the Monte-Carlo method.
We put the target velocity as $v_{\rm target} = 447$ m s$^{-1}$.
Since the maximum energy change with the sixty successive spin flipper units amounts $54$ $\mu$eV, the initial velocity of neutrons, which can be concentrated to the target velocity, distributes in the velocity region of $435$ to $447$ m s$^{-1}$.
Neutrons in this velocity range with unpreferred spin can be diluted into the wider velocity range of $410$ to $490$ m s$^{-1}$.
Thus we simulated the change of the velocity distribution in the velocity range of $410$ to $490$ m s$^{-1}$.
For simplicity, we assumed that the velocity distribution is flat and the time width of the neutron pulse is negligibly small.

\begin{figure}[t]
\centering\includegraphics[width=14.5cm]{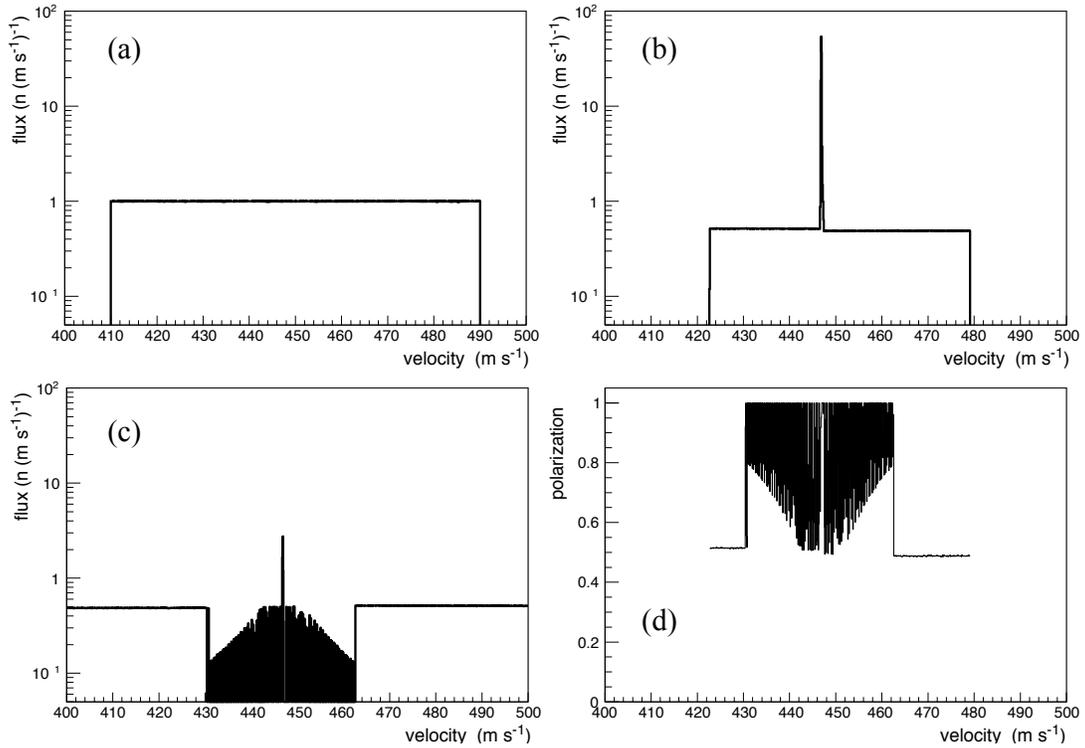}
\caption{Result of simulation of the velocity concentration.
(a) velocity distribution of flux of incident neutrons.
(b) exit flux of preferred spin component.
(c) exit flux of unpreferred spin component.
(d) polarization of exit neutrons. 
}
\label{fig_nvc1}
\end{figure}

Figure \ref{fig_nvc1} shows the result of the simulation.
Unpolarized neutrons in the velocity range shown in Fig.\ref{fig_nvc1} (a) was put in the initial condition.
As shown in Fig.\ref{fig_nvc1} (b), a sharp peak appears at the target velocity of $v_{\rm target}=447$ m s$^{-1}$ with the full width at the half maximum of $0.19$ m s$^{-1}$, which corresponds to the discrete energy change in individual spin flipper unit of $0.85$ $\mu$eV.
The neutron intensity is 50 times enhanced compared with the initial intensity.
Since the velocity concentrator is configured with the assembly of discrete length units, some of neutrons with unpreferred spin are also concentrated as shown in Fig. \ref{fig_nvc1} (c).
The resulting neutron polarization is almost 100\% as shown in Fig.\ref{fig_nvc1} (d).
In case polarized neutrons would be used, the net intensity enhancement is 100 in the vicinity of the target velocity.

\begin{figure}[th]
\centering\includegraphics[width=14.5cm]{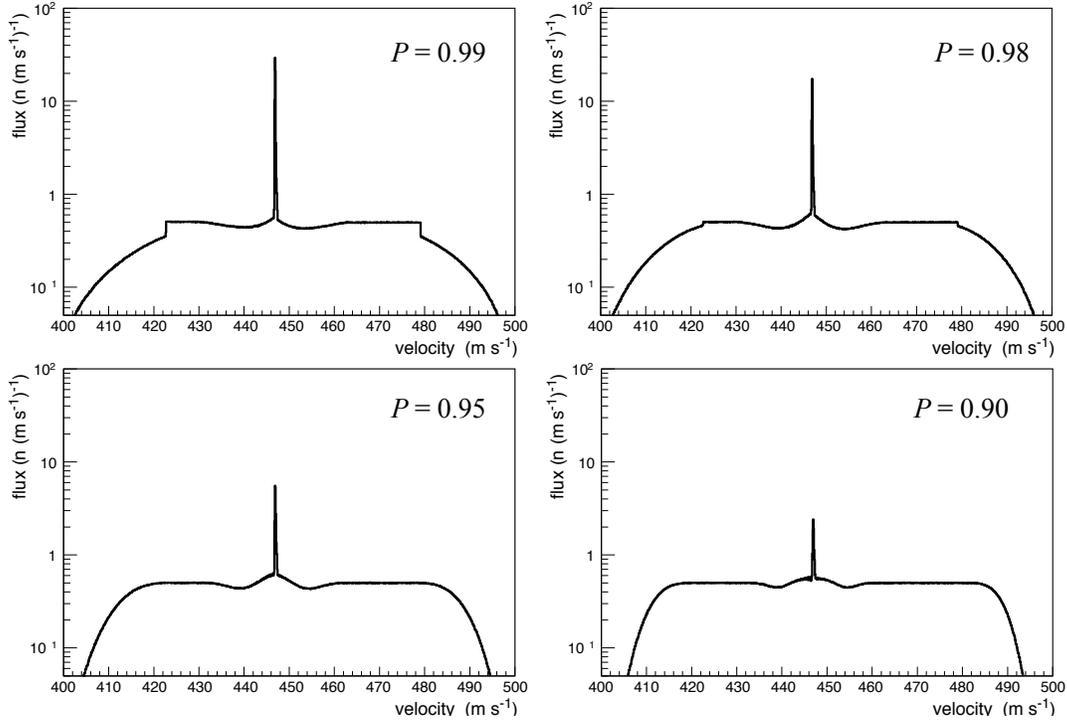}
\caption{The flux of exit neutrons from the incomplete flipping probability. 
$P$ is spin-flipping probability.}
\label{fig_nvc2}
\end{figure}

The spin flip probability was assumed to be 100\%  in the above result.
However, the practical flipper unit might have an incomplete spin flip probability.
The adverse effect of the incomplete spin flip probability was also simulated as shown in Fig.\ref{fig_nvc2}.
In these simulations, we ignored the quantum-mechanical phase relation between positive- and negative-polarity components and we  assumed that positive- and negative-polarity neutrons contributes incoherently.
The velocity concentration can be expected even with the spin flip efficiency as low as 90\% .

%

\section{Discussion}
The concept of the neutron velocity concentration was introduced with a possible design for the application to the enhancement of UCN production via the superthermal down-conversion in superfluid helium.
A possible enhancement with the velocity width of $0.19$ m s$^{-1}$ was estimated to be as large as 100.
We note that larger enhancement in narrower velocity width is achievable by employing additional spin flipper rf-cavities and/or frequency modulation synchronized with the neutron time-of-flight.
The optimum velocity width should be considered 

This technique can be applied to the nEDM measurement in the superfluid helium UCN converter located at the end of the neutron guide.
We note that the individual spin flipper units change the velocity distribution step by step and even smaller number of spin flipper units enables corresponding gain of UCN production.
We also note that the control of neutron energy with the high frequency spin flipper was successfully demonstrated~\cite{refRebuncher} and the fast switching of rf flippers synchronized with the neutron time-of-flight is already practically applied in J-PARC~\cite{refSFC}.
These achievements indicate that the neutron velocity concentration is feasible within the existing technologies.

%




\begin{thebibliography}{9}

\bibitem{refEDM}
C.A. Baker, et al., Phys. Rev. Lett. {\bf 97} 131801 (2006).

\bibitem{refLife}
A. P. Serebrov, Phys. Rev. C {\bf 82}, 035501 (2010).

\bibitem{refUCN}
R. Golub, D.J. Richardson, S.K. Lamoreaux, Ultracold Neutrons, Adam Hilger,
Bristol, 1991.

\bibitem{refANL}
A. Saunders, et al., Phys. Lett. B {\bf 593} 55 (2004).

\bibitem{refPSI}
A. Anghel, et al., Nucl. Instr. and Meth. A {\bf 611} 272 (2009).

\bibitem{refTRIUMF}
Y. Masuda, et al., Phys. Rev. Lett. {\bf 89} 284801 (2002).

\bibitem{refSNS}
N. Fomin, et. al., Nucl. Instr. and Meth. A {\bf 773}, 45-51 (2015).

\bibitem{refPNPI}
A. P. Serebrov, A.P. et al., Nucl.Instr. and Meth. A {\bf 611} 276 (2009).


\bibitem{refRebuncher}
Y. Arimoto, et al., Phys. Rev., A {\bf 86}, 023843 (2012).

\bibitem{refSFC}
K. Taketani, et. al., Nucl. Instr. and Meth. A {\bf 634} 134 (2011).

\end{thebibliography}
%

\end{document}